\documentclass{ws-ijmpcs}
\pdfoutput=1
\usepackage{cite}

\def\draftnote{}

\begin{document}

\markboth{S.\ H{\"o}che}
{Applications of higher order QCD}

%
\catchline{}{}{}{}{}
%

\title{Applications of higher order QCD}

\author{Stefan H{\"o}che}

\address{SLAC National Accelerator Laboratory, Menlo Park, CA 94025, USA\\
shoeche@slac.stanford.edu}

\maketitle
\hfill\begin{picture}(0,0)
  \put(-80,200){SLAC-PUB-15867}
\end{picture}


\begin{abstract}
In this talk we summarize some recent developments in perturbative QCD 
and their application to particle physics phenomenology.
\end{abstract}


\section{Introduction}
With the discovery of a new boson at the Large Hadron Collider (LHC), 
particle physics has entered a new era. Since this discovery, the field has quickly
moved towards precision measurements on the new particle. In order to further improve 
these measurements and to find possible small deviations that may hint towards new physics, 
improved theoretical predictions, including higher-order perturbative QCD corrections 
for production rates and kinematics are urgently needed. The same is true for 
other reactions of interest at the LHC, like top quark production and $W/Z$ production. 
The toolkit used to this end ranges from fixed order calculations at the parton-level 
over resummation to parton showers and particle-level event generators. 
Tremendous progress has been made in the field during the past year. 
Some of the recent developments will be briefly summarized in this talk.

\section{Higher-order calculations}
\label{sec:qcd_nlo}
Fixed-order calculations are available for a large variety of processes.
At the tree level, they have long been performed completely automatically 
using programs like ALPGEN \cite{Mangano:2002ea}, Amegic++ \cite{Krauss:2001iv}, 
Comix \cite{Gleisberg:2008fv}, CompHEP \cite{Boos:2004kh}, HELAC \cite{Kanaki:2000ey}, 
MadGraph \cite{Alwall:2011uj} and Whizard \cite{Kilian:2007gr}.
At the next-to-leading order (NLO), automation required two main ingredients: 
The implementation of known generic methods to perform the subtraction of 
infrared singularities \cite{Frixione:1995ms,Catani:1996vz,Catani:2002hc}, 
and the automated computation of one-loop amplitudes. As infrared subtraction terms 
consist of tree-level matrix elements joined by splitting operators, 
existing programs for leading order calculations are ideally suited 
to compute them. Correspondingly, Catani-Seymour dipole subtraction has
been implemented in the existing generators Amegic++ \cite{Gleisberg:2007md}, 
Comix, HELAC \cite{Czakon:2009ss} and MadGraph \cite{Frederix:2008hu,Frederix:2010cj}.
FKS subtraction is realized in MadGraph only \cite{Frederix:2009yq}.

\begin{figure}
  \begin{minipage}{0.475\textwidth}
    \includegraphics[width=\textwidth]{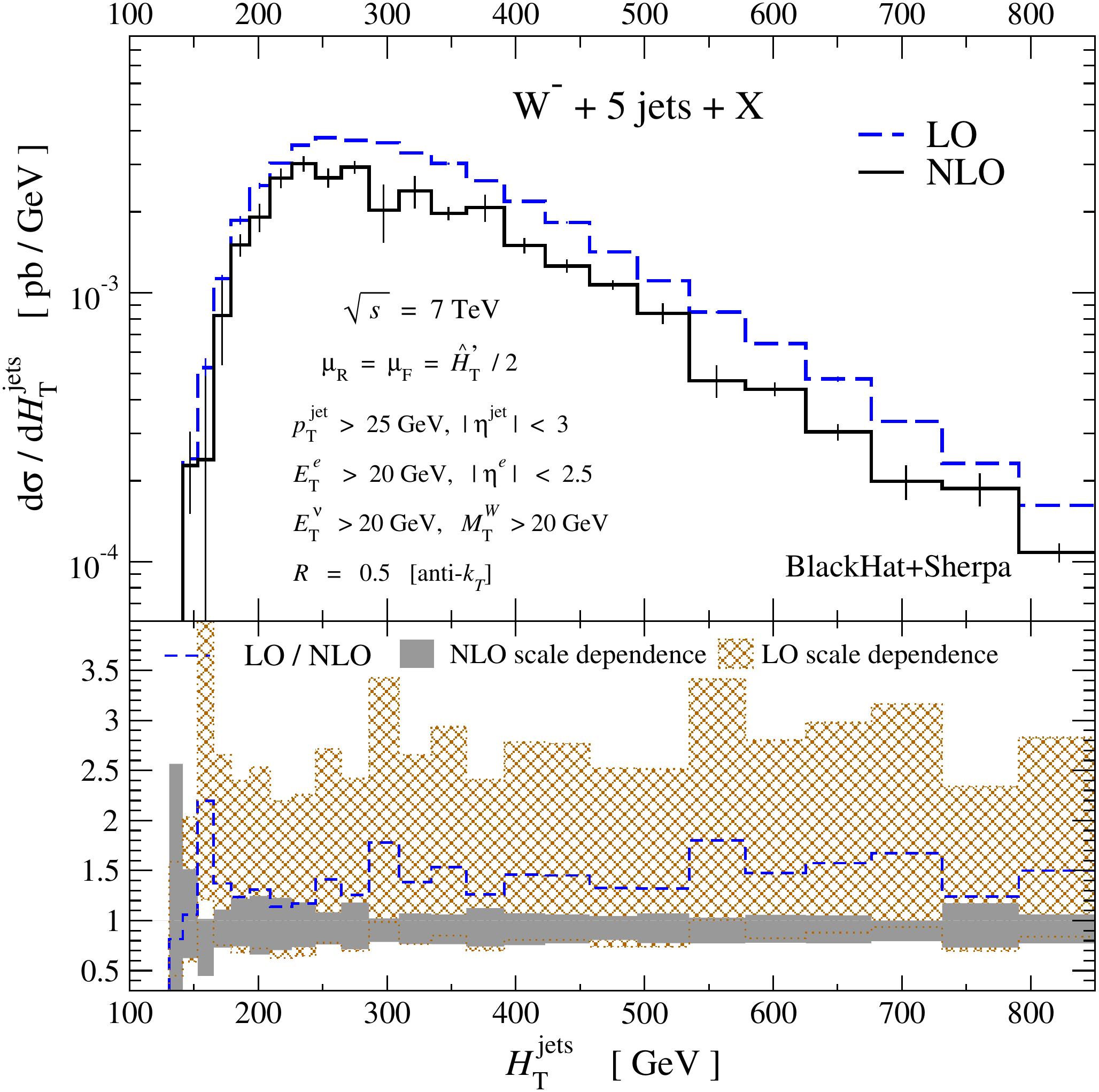}
    \caption{Distribution of the visible energy in $W+5$ jet events.
      Figure taken from \cite{Bern:2013gka}.
      \label{fig:qcd_w5jet}}
  \end{minipage}\hfill
  \begin{minipage}{0.475\textwidth}
    \includegraphics[width=\textwidth]{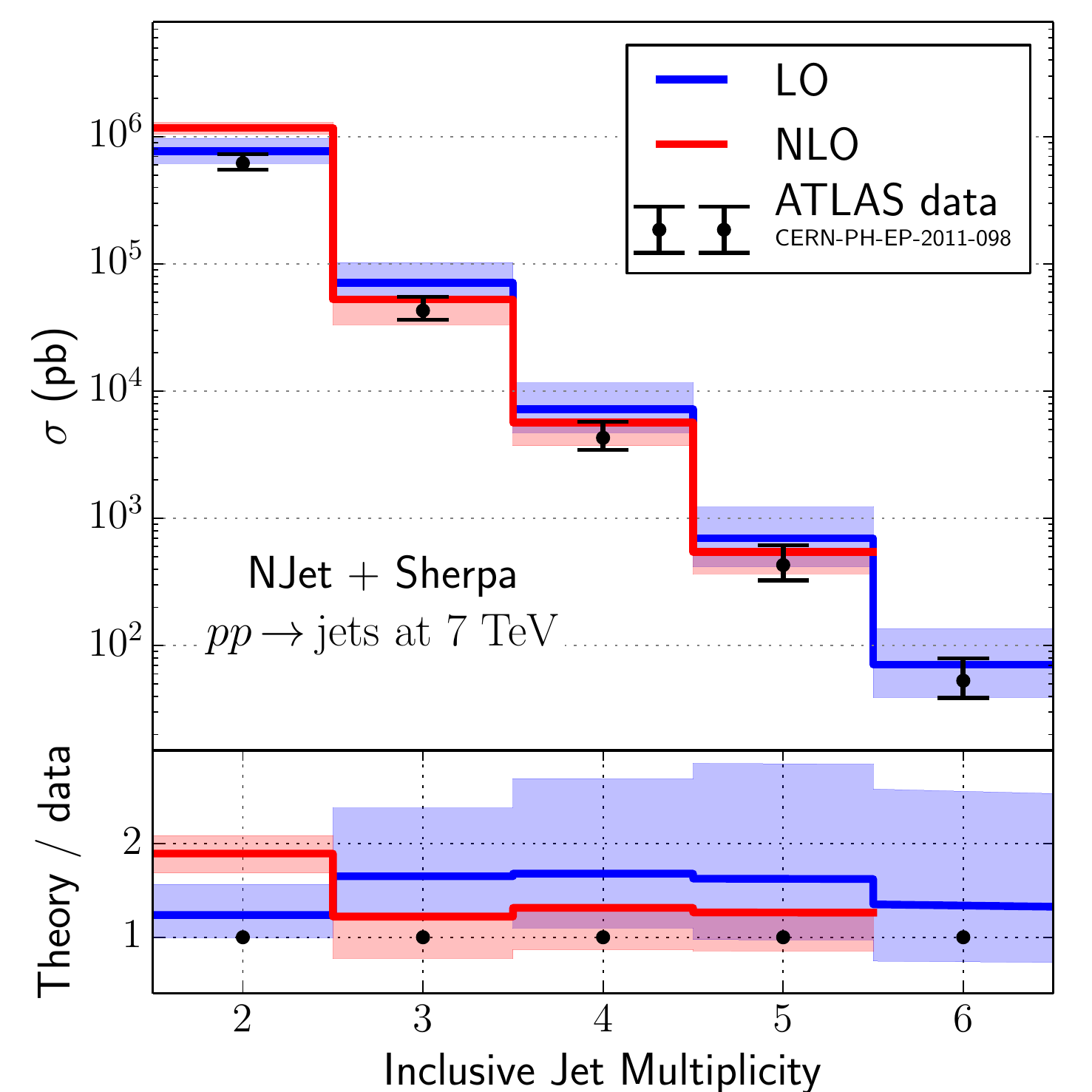}
    \caption{Jet multiplicity distribution in pure jet events (right).
      Figure taken from \cite{Badger:2013yda}.
      \label{fig:qcd_5jet}}
  \end{minipage}
\end{figure}

The automated computation of virtual corrections has received a boost 
from generalized unitarity \cite{Bern:1994cg,Bern:1994zx,Bern:1997sc}, 
which can be used to determine one-loop amplitudes by decomposing them into 
known scalar one-loop integrals and rational coefficients determined from 
tree amplitudes, plus a rational piece \cite{Ossola:2006us,Forde:2007mi,
  Ellis:2007br,Ossola:2008xq,Ellis:2008ir}. Programs like BlackHat 
\cite{Berger:2008sj}, Gosam \cite{Cullen:2011xs}, HELACNLO \cite{Bevilacqua:2011xh},
MadLoop \cite{Hirschi:2011pa}, NJet \cite{Badger:2010nx}, OpenLoops \cite{Cascioli:2011va} 
and Rocket \cite{Ellis:2008qc,Ellis:2009zw} implement these techniques and supplement 
established programs like MCFM \cite{MCFM,Campbell:2010ff} and dedicated codes 
based on improved tensor reduction approaches \cite{Denner:2005nn,Binoth:2005ff}. 
New techniques have also been proposed to accelerate the numerical calculation of the 
integrand of one-loop amplitudes, independent of the reduction scheme \cite{Cascioli:2011va}. 
Figures~\ref{fig:qcd_w5jet} and~\ref{fig:qcd_5jet} show examples from recent NLO calculations 
for $W$+5 jet production \cite{Bern:2013gka} and $5$ jets production \cite{Badger:2013yda},
both performed using unitarity based techniques. 
Other recently completed calculations include Higgs boson plus $3$ jet production 
\cite{Cullen:2013saa} and di-photon plus $2$ jet production \cite{Gehrmann:2013bga}.
The rapid progress in this field is reflected by the fact that
all calculations from the experimenter's wishlist for the LHC have now been tackled
\cite{AlcarazMaestre:2012vp}. Most of the programs used to perform the calculations, 
or their results, are publicly available.

\begin{figure}[t]
  \begin{minipage}{0.475\textwidth}
    \includegraphics[width=\textwidth]{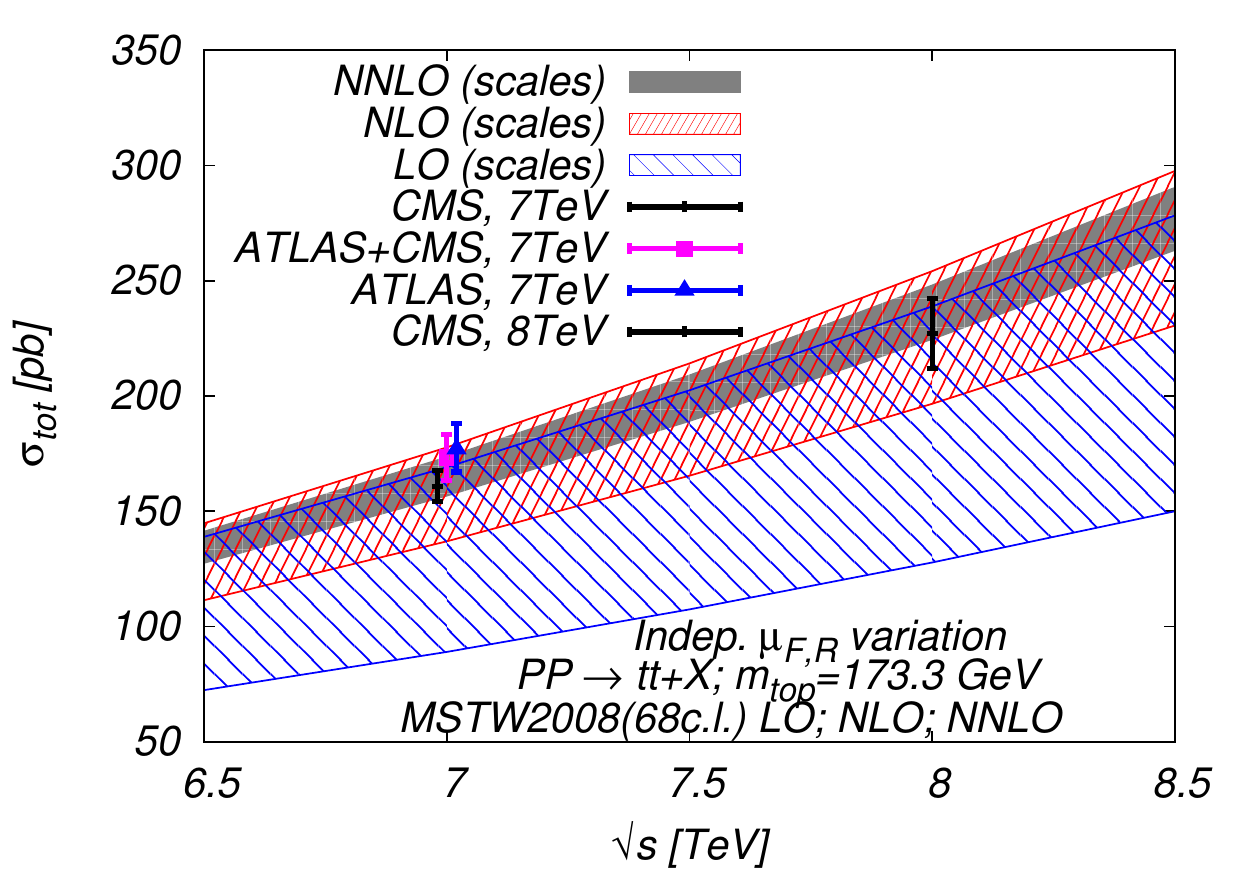}
    \caption{Energy dependence of the $pp\to t\bar{t}$ total cross.
      Figure taken from \cite{Czakon:2013goa}.
      \label{fig:qcd_tt}}
  \end{minipage}\hfill
  \begin{minipage}{0.475\textwidth}\vskip -2mm
    \includegraphics[width=\textwidth,clip,trim=3mm 0mm 6mm 0mm]{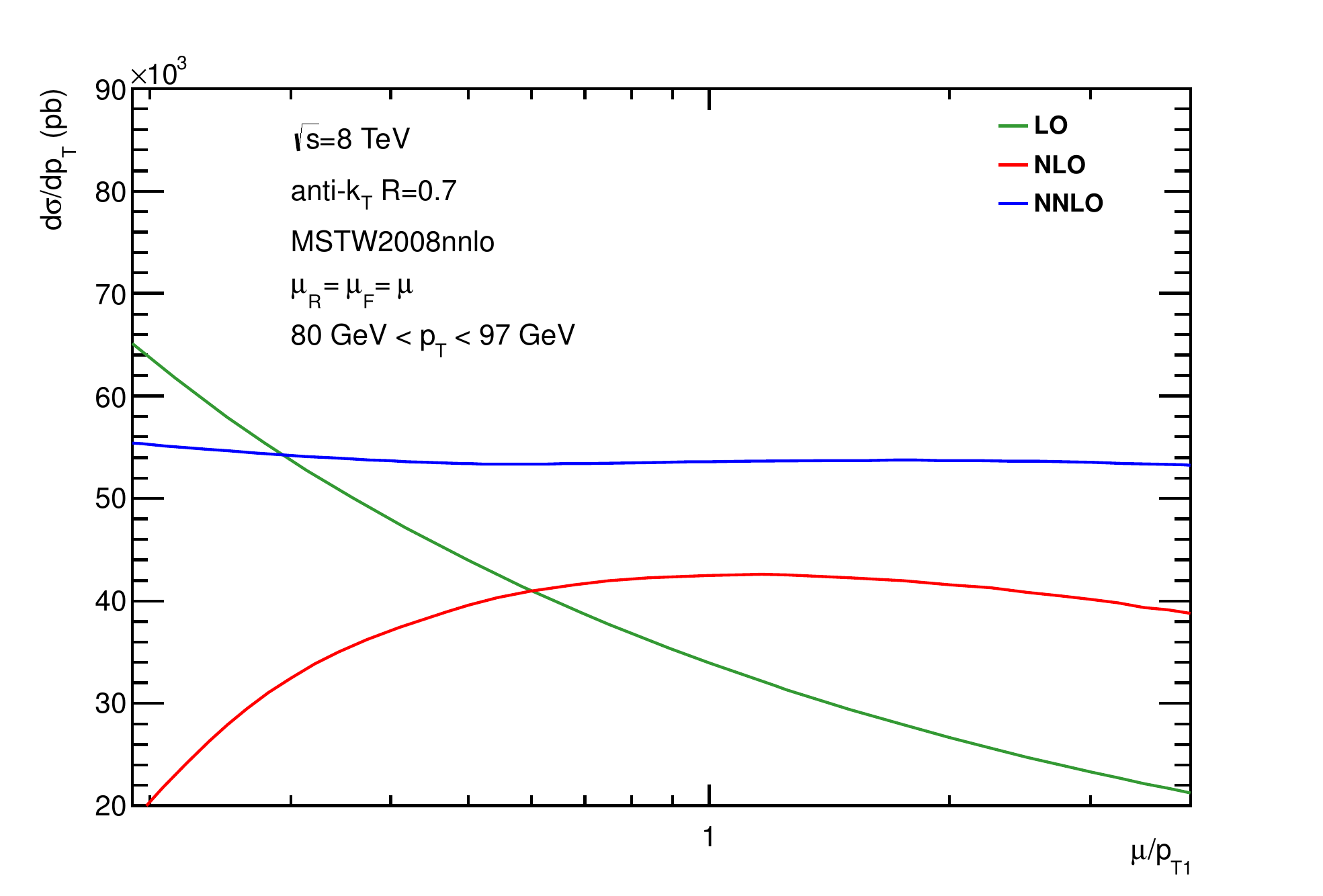}\vskip 2mm
    \caption{Scale dependence of the $pp\to$di-jet cross section.
      Figure taken from \cite{Ridder:2013mf}.
      \label{fig:qcd_jj}}
  \end{minipage}\hfill
\end{figure}

Driven by the need for higher precision in some selected Standard-Model reactions,
the field of next-to-next-to leading order (NNLO) calculations has significantly 
advanced in the past years. One of the most challenging problems is the regularization 
of infrared divergences at NNLO. Sector decomposition \cite{Binoth:2000ps,
  Anastasiou:2003gr,Binoth:2004jv}
has been used in the past to perform several $2\to 1$ calculations
\cite{Anastasiou:2004xq,Melnikov:2006di}.
Antenna subtraction \cite{Kosower:1997zr,GehrmannDeRidder:2005cm} 
was worked out and implemented for $e^+e^-\to 3$ jets 
\cite{GehrmannDeRidder:2007bj,GehrmannDeRidder:2007jk}. 
$q_T$ subtraction \cite{Catani:2007vq} was employed in several calculations, including
Higgs production \cite{Grazzini:2008tf}, $W/Z$ production \cite{Catani:2009sm}, 
associated Higgs production \cite{Ferrera:2011bk} and di-photon production \cite{Catani:2011qz}.
More recently sector-improved subtraction methods were introduced \cite{Czakon:2010td,Boughezal:2011jf}.
They have been used to compute cross sections for $pp\to t\bar{t}$
\cite{Baernreuther:2012ws,Czakon:2013goa} and $pp\to H+$jet \cite{Boughezal:2013uia}. 
At the same time, antenna subtraction was extended to 
initial states \cite{Daleo:2009yj,Boughezal:2010mc,Gehrmann:2011wi,GehrmannDeRidder:2012ja}
and employed to compute $pp\to$ di-jets fully differentially 
at NNLO \cite{Ridder:2013mf}. Figures~\ref{fig:qcd_tt} and \ref{fig:qcd_jj} show results from 
some of these calculations. The calculation of $pp\to t\bar{t}$ has also been combined with
higher logarithmic resummation \cite{Czakon:2009zw,Beneke:2009ye,Cacciari:2011hy}. 
Its theoretical uncertainty is such that uncertainties from scale choices, PDF, 
strong coupling measurements and top-quark mass measurements are all of the same order 
\cite{Czakon:2013tha}.

\section{Resummation of jet vetoes}
\label{sec:qcd_resum}
The analysis of the Higgs-like particle discovered at the LHC places new demands 
on resummed calculations.  Many of the Higgs analysis channels, most notably 
$H \to WW^* \to \ell^+ \ell^- \nu \bar{\nu}$, veto on the transverse momentum
of final state jets to distinguish different Standard Model backgrounds and 
separate them from the signal.  The leading systematic uncertainty is the
theoretical uncertainty on the signal cross section in the jet bins.
This uncertainty can be reduced by a proper resummation of the logarithms
associated with the jet veto. Various groups have investigated this problem, 
in most cases up to next-to-next-to leading logarithmic accuracy 
matched to NNLO fixed order, relying either on more traditional resummation methods 
\cite{Banfi:2012yh,Banfi:2012jm}, or on Soft Collinear Effective Theory 
\cite{Tackmann:2012bt,Stewart:2013faa,Becher:2012qa,Becher:2012yn,Becher:2013xia}.
Higgs plus one jet production was studied at next-to-leading logarithmic order
(NLL) and matched to NLO fixed order using SCET \cite{Liu:2013hba}.

\section{Parton showers and matching to NLO calculations}
\label{sec:qcd_meps}
The interest in parton showers as a means to produce particle-level predictions
fully differentially in the phase space of multi-jet events has increased significantly 
in recent years. New concepts for the construction of parton showers have been proposed,
which are based on antenna subtraction \cite{Giele:2007di,Hartgring:2013jma} 
and/or sectorizing the phase space \cite{Larkoski:2009ah,Larkoski:2011fd}.
Efforts were made to include subleading color corrections into showers as a means 
to improve their logarithmic accuracy \cite{Platzer:2012np,Hoeche:2011fd}.
However, the crucial development was the proposal of a method to match 
parton showers to NLO calculations \cite{Frixione:2002ik}, later extended 
to eliminate negative weights \cite{Nason:2004rx,Frixione:2007vw}.
This matching has been partially or fully automated in several projects 
\cite{Alioli:2010xd,Hoche:2010pf,Frederix:2011ig,Hoeche:2012ft,Hoeche:2012fm}, 
such that particle-level predictions at NLO accuracy are now widely available.

\begin{figure}[t]
  \begin{minipage}{0.47\textwidth}\vskip 4.5mm
    \includegraphics[width=\textwidth]{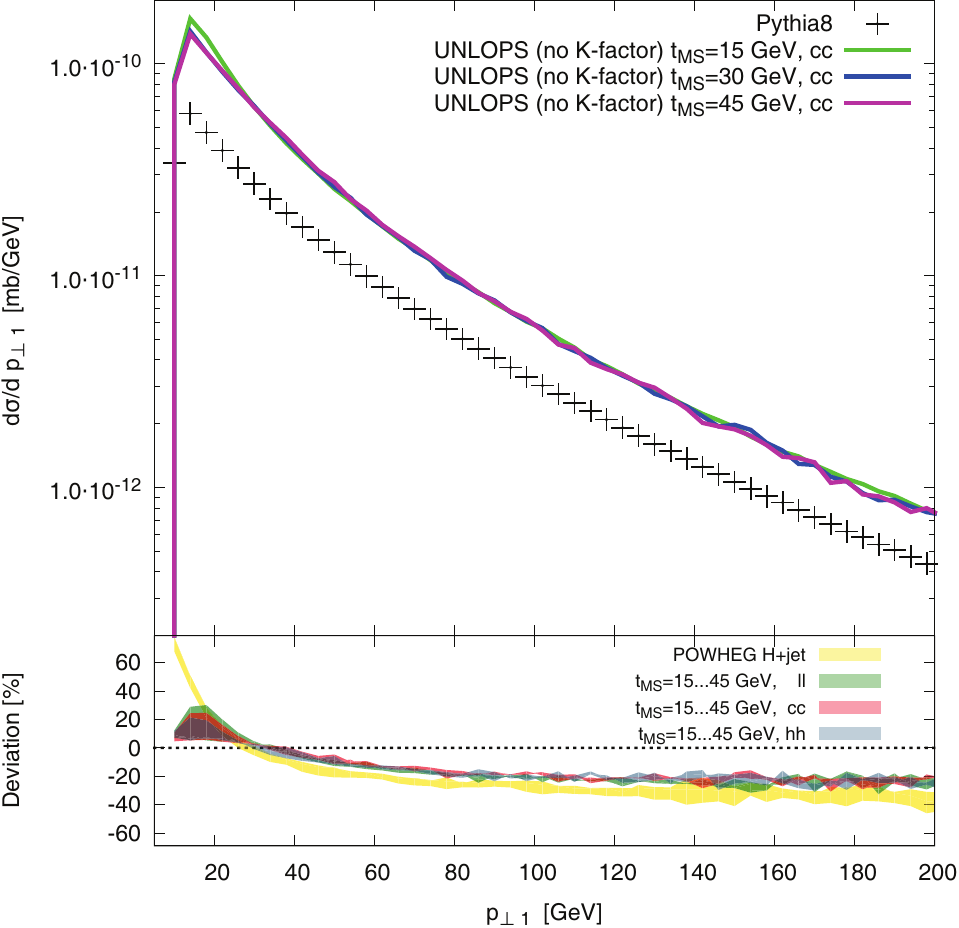}
    \caption{Transverse momentum of first jet in Higgs plus jets events.
      Figure taken from \cite{Lonnblad:2012ix}.
      \label{fig:qcd_meps_h}}
  \end{minipage}\hfill
  \begin{minipage}{0.48\textwidth}
    \includegraphics[width=\textwidth]{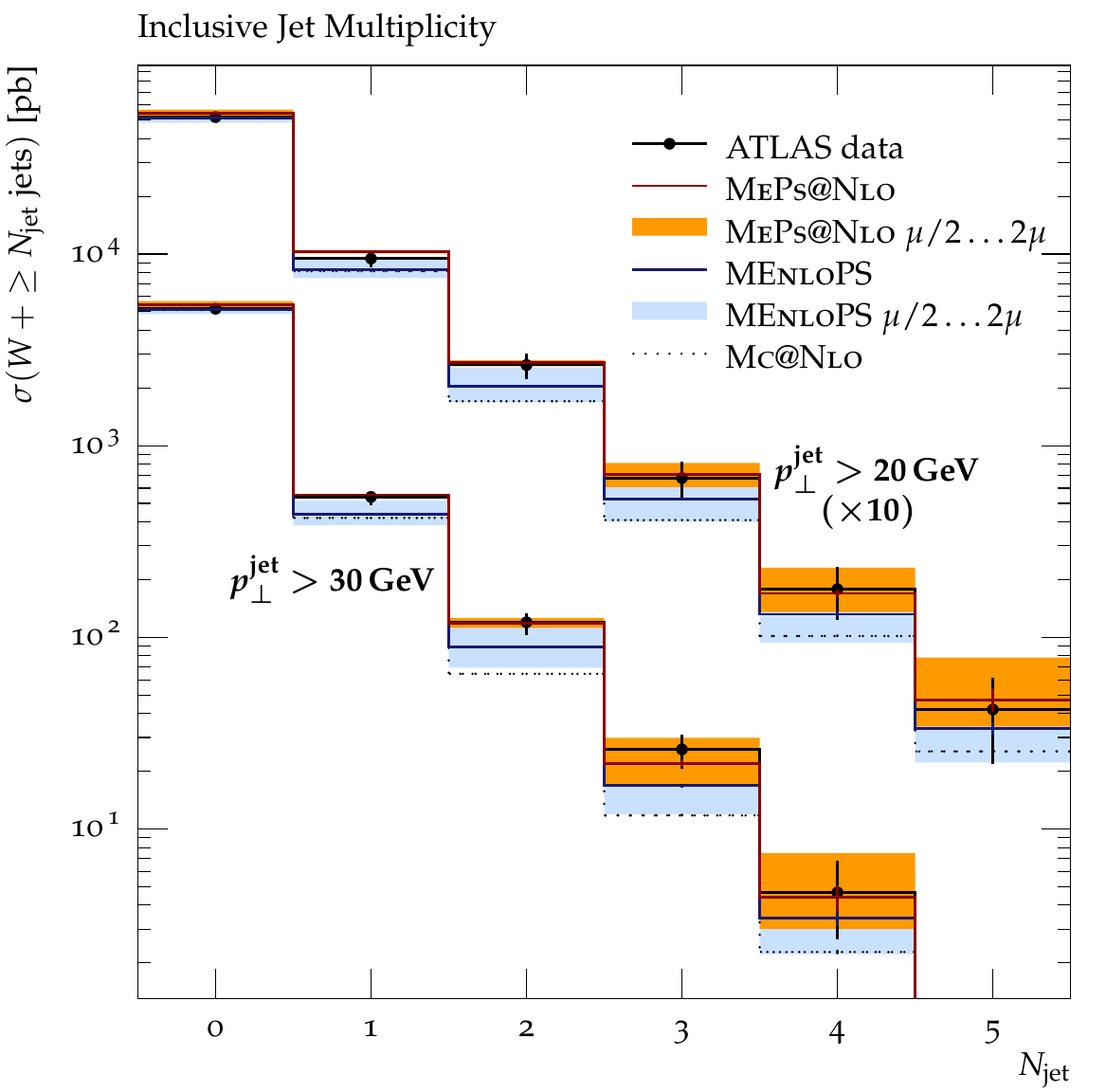}
    \caption{Jet multiplicity distribution in $W+$jets events.
      Figure taken from \cite{Hoeche:2012yf}.
      \label{fig:qcd_meps_w}}
  \end{minipage}\hfill
\end{figure}

The description of multi-jet final states with parton showers can be improved
using so-called ME+PS merging methods \cite{Catani:2001cc,Krauss:2002up,
  Mangano:2001xp,Alwall:2007fs,Lonnblad:2012ng}, 
which, in contrast to matching methods, allow to correct the parton shower 
for an arbitrary number of emissions with higher-order tree-level calculations. 
These methods were recently refined and extended,
leading to algorithms which can combine multiple NLO calculations of varying 
multiplicity (like $W+0$ jet, $W+1$ jet, $W+2$ jet, etc.) into a single, inclusive
simulation (e.g.\ of $W+$jets production) \cite{Lavesson:2008ah,Gehrmann:2012yg,
  Hoeche:2012yf,Lonnblad:2012ix,Frederix:2012ps}. Figures~\ref{fig:qcd_meps_h}
and \ref{fig:qcd_meps_w} show examples for the application of ME+PS merging to 
Higgs boson plus jets production and to $W+$jets production.
A particular scale choice is required for the evaluation of the strong coupling
in ME+PS  merging, which has also been adopted for the matching to higher-multiplicity 
NLO calculations on its own in the so-called MINLO approach \cite{Hamilton:2012np}. 

The MINLO method accounts for Sudakov suppression effects in higher-multiplicity
final states and allows to extrapolate NLO calculations to zero jet transverse 
momentum, thus offering the opportunity to match to NNLO calculations for a limited
class of processes and observables \cite{Hamilton:2012rf}. 
A different proposal for a matching to NNLO parton-level calculations was made
in \cite{Lavesson:2008ah,Lonnblad:2012ix}, which is based on a subtraction method 
similar to the one used in ME+PS merging at NLO. Both techniques are promising candidates 
to further increase the precision of event generators for collider physics.

\section{Summary}
We have presented some of the recent developments in perturbative QCD and
applications to particle physics phenomenology. NLO parton-level calculations 
can nowadays often be provided by fully automated tools. 
New techniques in event generation allow to also use them for particle-level
predictions. NNLO calculations and higher-logarithmic resummation techniques
are at the forefront of current research.

\bibliographystyle{ws-ijmpcs}  
\bibliography{journal}

\end{document}